\pdfoutput=1
\RequirePackage{fix-cm}
\documentclass[smallextended]{svjour3}       
\smartqed  
\usepackage{graphicx}
\usepackage{graphicx}
\usepackage{amsmath}
\usepackage{multirow}
\usepackage{booktabs}
\begin{document}

\title{Twitter Sentiment Analysis: How To Hedge Your Bets In The Stock Markets}
\subtitle{}


\author{Tushar Rao         \and
        Saket Srivastava 
}


\institute{T. Rao \at
              Netaji Subhas Institute of Technology, Delhi, India \\
              \email{rao.tushar@nsitonline.in}           
           \and
           S. Srivastava \at
              Indraprastha Institute of Technology, Delhi, India\\
              \email{saket@iiitd.ac.in}\\
}
\date{Received: 5th December 2012}
\maketitle

\begin{abstract}
Emerging interest of trading companies and hedge funds in mining social web has created new avenues for intelligent systems that make use of public opinion in driving investment decisions. It is well accepted that at high frequency trading, investors are tracking memes rising up in microblogging forums to count for the public behavior as an important feature while making short term investment decisions. We investigate the complex relationship
between tweet board literature (like bullishness, volume, agreement etc)
with the financial market instruments (like volatility, trading
volume and stock prices). We have analyzed
Twitter sentiments for more than 4 million tweets between
June 2010 and July 2011 for DJIA, NASDAQ-100 and 11 other big cap technological stocks. Our results show high correlation
(upto 0.88 for returns) between stock prices and twitter sentiments.
Further, using Granger's Causality Analysis, we have validated
that the movement of stock prices and indices are greatly affected
in the short term by Twitter discussions. Finally, we have implemented
Expert Model Mining System (EMMS) to demonstrate that our forecasted
returns give a high value of R-square (0.952) with low Maximum Absolute
Percentage Error (MaxAPE) of 1.76\% for Dow Jones Industrial Average (DJIA). We introduce a novel way to make use of market monitoring elements derived from public mood to retain a portfolio within limited risk state (highly improved hedging bets) during typical market conditions.

\end{abstract}

\section{INTRODUCTION}
\label{intro}
Financial analysis and computational finance have been an active
area of research for many decades\cite{Lee1999357}. Over the years, several new
tools and methodologies have been developed that aim to predict
the direction as well as range of financial market instruments as
accurately as possible\cite{Guresen201110389}. Before the emergence of
internet, information regarding company's stock price, direction
and general sentiments took a long time to disseminate among
people. Also, the companies and markets took a long time (weeks or months)
to calm market rumors, news or false information (memes in Twitter context).
Web $3.0$ is characterized with fast pace information
dissemination as well as retrieval~\cite{danah_article}. Spreading good or bad
information regarding a particular company, product, person etc.
can be done at the click of a mouse~\cite{Brown:2002:SLI:560498}, \cite{Acemoglu2010194}
or even using micro-blogging services such as Twitter\cite{BBC_twitter}.
Recently scholars have made use of twitter feeds in predicting box
office revenues~\cite{Asur_Huberman_2010}, political game
wagons~\cite{Tumasjan_Sprenger_Sandner_Welpe_2010},
rate of flu spread~\cite{Szomszor_Kostkova_Quincey_2009} and
disaster news spread~\cite{tweet_disaster}. For short term trading decisions, {\it short term sentiments}
play a very important role in {\it short term performance}
of financial market instruments such as indexes, stocks and bonds~\cite{morning_paper}.

Early works on stock market prediction can be
summarized to answer the question - Can stock prices be really
predicted? There are two theories - (1) random walk theory (2) and efficient market hypothesis
(EMH)\cite{Hong}. According to EMH stock index largely
reflect the already existing news in the investor community rather
than present and past prices. On the other hand, random walk theory
argues that the prediction can never be accurate since the time instance
of news is unpredictable. A research conducted by Qian et.al.
compared and summarized several theories that challenge the basics
of EMH as well as the random walk model completely\cite{Qian}. Based
on these theories, it has been proven that some level of prediction
is possible based on various economic and commercial indicators.
The widely accepted semi-strong version of the EMH claims that prices aggregate all publicly
available information and instantly reflect new public version\cite{Burton_m}.
It is well accepted that {\it news drive macro-economic movement in the markets}, while researches suggests
that {\it social media buzz is highly influential at micro-economic level}, specially
in the big indices like DJIA \cite{Bollen_Mao_Zeng_2010}, \cite{Bollen_second_paper}, \cite{Gilbert_Karahalios_2010} and \cite{Sprenger}. Through earlier researches it has been validated that market is
completely driven by sentiments and bullishness of the investor's
decisions \cite{Qian}. Thus a comprehensive model that could incorporate these
sentiments as a parameter is bound to give superior prediction at {\it micro-economic}
level.

Earlier work done by Bollen et. al. shows how collective
mood on Twitter (aggregate of all positive and
negative tweets) is reflected in the DJIA index movements~\cite{Bollen_Mao_Zeng_2010}and \cite{Bollen_second_paper}.
In this work we have applied simplistic message board approach by defining bullishness and agreement terminologies derived from positive and negative vector ends of public sentiment w.r.t. each market security or index terms (such as returns, trading volume and volatility). Proposed method is not only scalable but also gives more accurate measure of large scale investor sentiment that can be potentially used for short term hedging strategies as discussed ahead in section \ref{hedging_startgies}.
This gives clear distinctive way for modeling sentiments for service based companies such as Google in contrast to product based companies such as Ebay, Amazon and Netflix. We validate that Twitter feed for any company reflects
the public mood dynamics comprising of breaking news and discussions, which is causative in nature. Therefore it adversely
affects any investment related decisions which are not limited to stock
discussions or profile of mood states of entire Twitter feed.

In section~\ref{motivation}, we discuss the motivation of this
work and related work in the area of stock market prediction
in section~\ref{related_work}. In section~\ref{data_collection} we
explain\textit{ what and how} of the techniques used in mining data and
explain the terminologies used in market and tweet board
literature. In section~\ref{results} we have given prediction methods
used in this model with the forecasting results. In section \ref{hedging_startgies} we discuss how Twitter based model can be used for improving hedging decisions in a diversified portfolio by any trader.
Finally in section \ref{discussions} we discuss the results and in section \ref{conclusion} we present the future prospects and
conclude the work.

\section{MOTIVATION}
\label{motivation}

\textit{"Communities of active investors and day traders who are
sharing opinions and in some case sophisticated research about stocks,
bonds and other financial instruments will actually have the power to
move share prices ...making Twitter-based input as important as any
other data to the stock"}
\begin{flushright}
     -TIME (2009) \cite{TIME09}
\end{flushright}

\begin {figure*}[htll]
\centering
\includegraphics [scale=0.4]{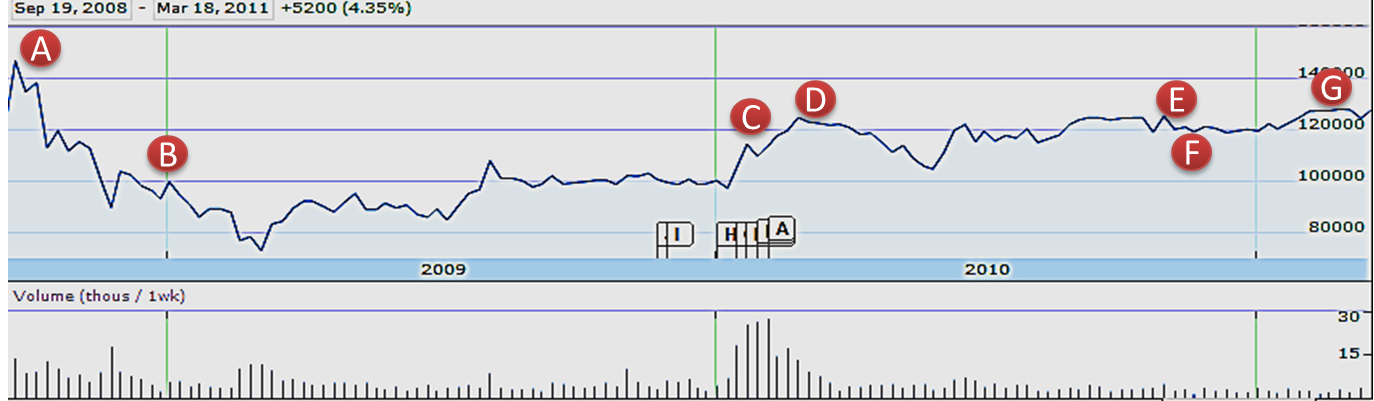}
\caption {Historical chart of Berkshire Hathaway(BRK.A) stock over the last 3 years. Highlighted points (A-F) are the days when its stock price jumped due to an increased news volume on social networks and Twitter regarding Anne Hathaway. Courtesy Google Finance.}
\label{fig:Hathaway}
\end{figure*}

High Frequency Trading (HFT) comprises of very high percentage of trading volumes in the present US market. Traders make an investment position that is held only for very brief periods of time - even just seconds - and rapidly trades into and out of those positions, sometimes thousands or tens of thousands of times a day. Therefore the value of an investment
is as good as last known index price. Investors will do anything
that will give them an advantage in placing market bets. A large
percentage of high frequency traders in US markets, have
trained AI bots to capture buzzing trends in the social media
feeds without learning dynamics of the sentiment and accurate
context of the deeper information being diffused in
the social networks. For example, in February 2011 during Oscars when Anne Hathaway was trending,
stock prices of Berkshire Hathaway rose by 2.94\% \cite{Hathaway}
Figure~\ref{fig:Hathaway} highlight the incidents when the stock price of Berkshire Hathaway
jumped coinciding with an increase of buzz on social networks/ micro-blogging websites
regarding Anne Hathaway (for example during movie releases).

\textsl{The events are marked as red points in the Figure \ref{fig:Hathaway} , event specific news on the points- \\
A: Oct. 3, 2008 - Rachel Getting Married opens: BRK.A up 0.44\%\\
B: Jan. 5, 2009 - Bride Wars opens: BRK.A up 2.61\%\\
C: Feb. 8, 2010 - Valentine's Day opens: BRK.A up 1.01\% \\
D: March 5, 2010 - Alice in Wonderland opens: BRK.A up 0.74\% \\
E: Nov. 24, 2010 - Love and Other Drugs opens: BRK.A up 1.62\% \\
F: Nov. 29, 2010 - Anne announced as co-host of the Oscars: BRK.A up 0.25\% \\
G: Feb. 28. 2011 - Anne hosts Oscars with James Franco: BRK.A up 2.94\% \\}

As seen in this example, large volume of tweets can create~\textit{short term influential effects} on
stock prices. Events such as these motivate us to investigate deeper relationship
between the dynamics of social media messages and market movements~\cite{Lee1999357}. This work is not directed to find a new stock prediction technique
which will counter in the effects of various other macroeconomic
factors.

{\it The aim of this work}, is to quantitatively evaluate the {\it effects
of twitter sentiment dynamics} around a stocks indices/stock prices and use
it in conjunction with the {\it standard} model to improve the
accuracy of prediction.  Further in section \ref{hedging_startgies} we investigate into how tweets can be very useful in identifying trends in futures
and options markets and to build hedging strategies to protect
one's investment position in the shorter term.

\begin {figure*}[htll]
\centering
\includegraphics [scale=0.8] {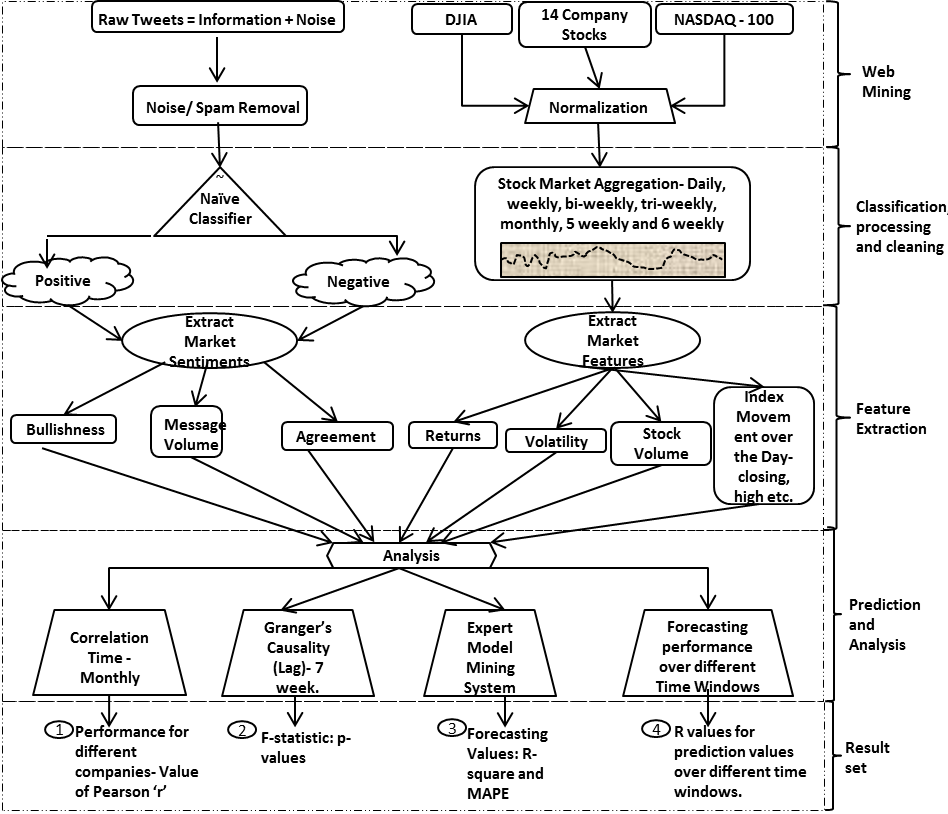}
\caption {Flowchart of the proposed methodology showing the various phases
of sentimental analysis beginning with Tweet collection to stock future prediction.
In the final phase 4 set of results have been presented:(1) Correlation
results for twitter sentiments and stock prices for different companies
(2) Granger's casuality analysis to prove that the stock prices are affected in
 the short term by Twitter sentiments (3) Using EMMS for quantitative
 comparison in stock market prediction using tweet features
(4) Performance of Twitter sentiment forecasting method over different time windows}
\label{fig:Work Process}
\end{figure*}

\section{RELATED WORK}
\label{related_work}

There have been several works related to web mining of data
(blogposts, discussion boards and news)~\cite{Ant_frank}, \cite{Bagnoli199927}, \cite{Gilbert_Karahalios_2010}
and to validate the significance of assessing behavioral changes
in the public mood to track movements in stock markets. Some trivial work shows
information from investor communities is causative of speculation
regarding private and forthcoming information and commentaries\cite{lerman}, \cite{Wysocki_1998},\cite{Das_chen} and \cite{Da_Engelberg_Gao_2010}.
Dewally in 2003 worked upon naive momentum strategy confirming
recommended stocks through user ratings had significant prior
performance in returns \cite{Dewally}. But now with the pragmatic shift in the online habits of communities around the worlds, platforms like StockTwits\footnote{http://stocktwits.com/}~\cite{Business_week} and
TweetTrader\footnote{http://tweettrader.net/} have come up and their usage is virally spreading
out.
Das and Chen made the initial attempts by using natural language
processing algorithms classifying stock messages based on human
trained samples. However their result did not carried
statistically significant predictive relationships \cite{Das_chen}.

Gilbert et.al. and Zhang et.al. have used
corpus from livejournal blogposts in assessing the bloggers
sentiment in dimensions of fear , anxiety and worry making use of
Monte Carlo simulation to reflect market movements in S\&P 500
index \cite{Gilbert_Karahalios_2010,Zhang_Fuehres_Gloor_2009}. Similar and significantly accurate work is done by
Bollen et. al who used dimensions of Google- Profile of Mood
States to reflect changes in closing price of DJIA \cite{Bollen_Mao_Zeng_2010}. Sprengers et.al. analyzed individual stocks for S\&P 100 companies and tried
correlating tweet features about discussions of the
stock discussions about the particular companies containing the
Ticker symbol \cite{Sprenger}.
However these approaches have been
restricted to community sentiment at macro-economic level which doesn't give explanatory dynamic system for
individual stock index for companies. Thus deriving a model that is scalable for individual stocks/ companies and can
be exploited to make successful hedging strategies as discussed in section \ref{hedging_startgies}.
\section {WEB MINING AND DATA PROCESSING}
\label{data_collection}
In this section we describe our method of Twitter and financial
data collection as shown in Figure~\ref{fig:Work Process}.
In the first phase, we mine the tweet data and after removal
of spam/noisy tweets, they are subsequently subjected to
sentiment assessment tools in phase two. In later phases
feature extraction, aggregation and analysis is done.

\subsection {Tweets Collection and Processing}

Out of other investor forums and discussion boards, Twitter
has widest acceptance in the financial community and all
the messages are accessible through a simple search of
requisite terms through an application programming
interface (API)\footnote {Twitter API is easily accessible
through an easy documentation available at- https://dev.twitter.com/docs.
Also Gnip - http://gnip.com/twitter, the premium platform
available for purchasing public firehose of tweets has
many investors as financial customers researching in the
area.}. Sub forums of Twitter like StockTwits
and TweetTrader have emerged recently as hottest place
for investor discussion buy/sell out at voluminous rate.
Efficient mining of sentiment aggregated around these tweet
feeds provides us an opportunity to trace out relationships
happening around these market sentiment terminologies.
Currently more than 250 million messages are posted
on Twitter everyday (Techcrunch October 2011\footnote{http://techcrunch.com/2011/10/17/twitter-is-at-250-million-tweets-per-day/}).

This study was conducted over a period of 14 months period
between June 2nd 2010 to 29th July 2011. During this period,
we collected 4,025,595 (by around 1.08M users)
English language tweets Each tweet record contains (a) tweet identifier,
(b) date/time of submission(in GMT), (c) language and (d)text.
Subsequently the stop words and punctuation are removed and
the tweets are grouped for each day (which is the highest
time precision window in this study since we do not group
tweets further based on hours/minutes). We have
directed our focus DJIA, NASDAQ-100 and 11 major companies
listed in Table~\ref{tab:Company_list}. These companies are
some of the highly traded and discussed technology stocks
having very high tweet volumes.

\begin{table}[h!]
  \centering
  \caption{List of Companies}
  \scriptsize
    \begin{tabular}{ll}
    \toprule
    \textbf{Company Name} & \textbf{Ticker Symbol} \\
    \midrule
    \textbf{Amazon} & AMZN \\
    \textbf{Apple} & AAPL \\
    \textbf{AT\&T} & T \\
    \textbf{Dell} & DELL \\
    \textbf{EBay} & EBAY \\
    \textbf{Google} & GOOG \\
    \textbf{Microsoft} & MSFT \\
    \textbf{Oracle} & ORCL \\
    \textbf{Samsung Electronics} & SSNLF \\
    \textbf{SAP} & SAP \\
    \textbf{Yahoo} & YHOO \\
    \bottomrule
    \end{tabular}%
  \label{tab:Company_list}%
\end{table}%

\begin {figure}[htll]
\centering
\includegraphics [scale=0.58]{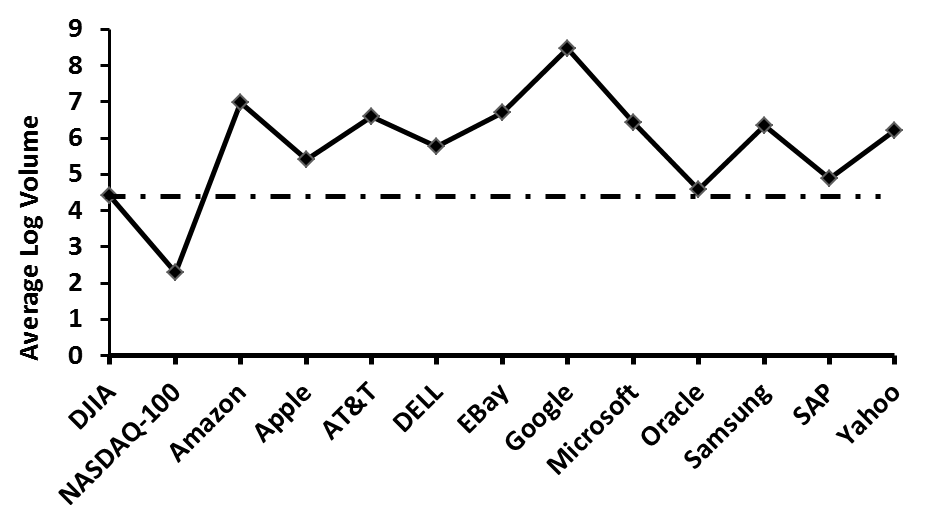}
\caption {Graph for average of log of daily volume over the months under study}
\label{fig:volume_of stocks}
\end{figure}

As seen in Figure~\ref{fig:volume_of stocks}, the average message
volume for the 11 companies used to validate the working model;
is more than the average discussion volume of DJIA and NASDAQ-100.
In this study we have observed that technology stocks generally
have a higher tweet volume than non-technology stocks. One reason
for this may be that all technology companies
come out with new products and announcements much more frequently
than companies in other sectors(say infrastructure, energy, FMCG, etc.)
thereby generating greater buzz on social media networks. However, our
model may be applied to any company/indices that generate high tweet volume.

\subsection {Sentiment Classification}
In order to compute sentiment for any tweet we had to classify each
incoming tweet everyday into {\it positive} or {\it negative} using naïve classifier. For each day total number of positive tweets is aggregated as $Positive_{day}$ while total number of negative tweets as $Negative_{day}$. We have made use of JSON API from Twittersentiment \footnote{https://sites.google.com/site/twittersentimenthelp/},
a service provided by Stanford NLP research group \cite{Alec_Bhayani}.
Online classifier has made use of Naive Bayesian classification method,
which is one of the successful and highly researched algorithms
for classification giving superior performance to other methods
in context of tweets. Their classification training was done over a dataset of 1,600,000 tweets and
achieved an accuracy of about 82.7\%. These methods have high replicability and few
arbitrary fine tuning elements.

In our dataset roughly 61.68\% of the tweets are positive,
while 38.32\% of the tweets are negative for the company
stocks under study. The ratio of 3:2 indicates stock discussions to be
much more balanced in terms of bullishness than internet board
messages where the ratio of positive to negative ranges from
7:1~\cite{Dewally} to 5:1~\cite {Ant_frank}.
Balanced distribution of stock discussion provides us with
more confidence to study information content of the positive and
negative dimensions of discussion about the stock prices on
microblogs.
\subsection {Tweet Feature Extraction}

One of the research questions this study explores is how
investment decisions for technological stocks are affected by entropy of information
spread about companies under study in the virtual space. Tweet messages
are micro-economic factors that affect stock prices which is quite
different type of relationship than factors like news aggregates from traditional
media, chatboard room etc. which are covered in earlier studies
over a particular period \cite{Dewally}, \cite{lerman}and  \cite{Ant_frank}.
Keeping this in mind we have only aggregated the tweet parameters
(extracted from tweet features) over a day.
In order to calculate parameters weekly, bi-weekly, tri-weekly,
monthly, 5 weekly and 6 weekly we have simply taken average of
daily twitter feeds over the requisite period of time.

\begin {figure}[hcc]
\centering
\includegraphics [scale=0.45]{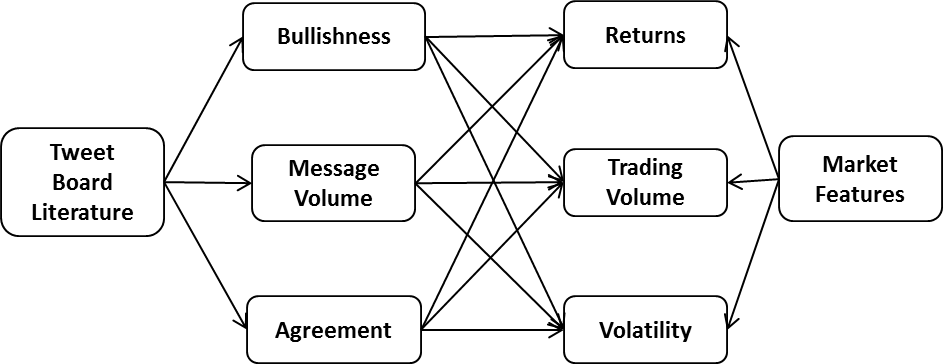}
\caption{Tweet Sentiment and Market Features}
\label{fig:Market_board}
\end{figure}

Twitter literature in perspective of stock investment
is summarized in Figure~\ref{fig:Market_board}. We have
carried forward work of Antweiler et.al. for
defining bullishness ($B_t$) for each day (or time window)
given equation as:

\begin{equation}\label{bullishness}
    B_t = \ln\left({\frac{1+{M_t}^{Positive}}{1+{M_t}^{Negative}}}\right)
\end{equation}

Where ${M_t}^{Positive}$ and ${M_t}^{Negative}$ represent number of positive
or negative tweets on a particular day $t$. Logarithm of bullishness measures
the share of surplus positive signals and also gives more weight to larger
number of messages in a specific sentiment (positive or negative).
Message volume  for a time interval \textit{t} is simply defined as natural logarithm of total number of tweets for a specific stock/index which is $ \ln ({M_t}^{Positive}+ {M_t}^{Negative})$. The agreement among positive and negative
tweet messages is given by:

\begin {equation}
A_t = 1- \sqrt{1- \frac{M_t^{Positive}-
M_t^{Negative}}{M_t^{Positive}+ M_t^{Negative}}}
\label{agreement}
\end {equation}

If $all$ tweet messages about a particular company are bullish or bearish,
agreement would be $1$ in that case. Influence of silent tweets days
in our study (trading days when no tweeting happens about
particular company) is less than $0.1\%$ which is significantly less
than previous research~\cite{Ant_frank,Sprenger}. Carried terminologies
for all the tweet features\{Positive, Negative, Bullishness, Message Volume, Agreement\}
remain same for each day with the lag of one day. For example, carried bullishness
for day $d$ is given by $Carried Bullishness _{d-1}$.\\

\subsection{Financial Data Collection}

We have downloaded financial stock prices at daily intervals from
Yahoo Finance API\footnote{http://finance.yahoo.com/} for DJIA, NASDAQ-100
and the companies under study given in Table~\ref{tab:Company_list}.
The financial features (parameters) under study are opening ($O_t$) and closing
($C_t$) value of the stock/index, highest ($H_t$) and lowest ($L_t$) value of
the stock/index and returns. Returns are calculated as
the difference of logarithm to the base $e$ between the closing values of the stock price of
a particular day and the previous day.

\begin {equation}
 R_t= \{{\ln Close_{(t)}-\ln Close_{(t-1)}}\}\times100
\end {equation}

Trading volume is the logarithm of number of traded shares. We estimate
daily volatility based on intra-day highs and lows using Garman and
Klass volatility measures~\cite{Garman_Klass} given by
the formula:

\begin{equation}
\sigma= \sqrt{\frac{1}{n}\sum{\frac{1}{2}[\ln{\frac{H_t}{L_t}}]^2
- [2\ln{2}-1][\ln{\frac{C_t}{O_t}}]^2}}
\end {equation}

\section{STATISTICAL ANALYSIS AND RESULTS}
\label{results}

We begin our study by identifying the correlation between
the Twitter feed features and stock/index parameters which
give the encouraging values of statistically significant
relationships with respect to individual stocks(indices).
To validate the causative effect of tweet feeds on stock
movements we have used econometric technique of Granger's
Casuality Analysis. Furthermore, we make use of expert model
mining system (EMMS) to propose an efficient prediction
model for closing price of DJIA and NASDAQ $100$. Since
this model does not allow us to draw conclusion about the accuracy
of prediction (which will differ across size of the
time window) subsequently discussed later in this section.

\subsection{Correlation Matrix}
For the stock indices DJIA and NASDAQ and 11 tech companies under study
we have come up with the correlation matrix given in Figure \ref{fig:corr_heatmap} in the appendix
between the financial market and Twitter sentiment features explained in
section~\ref{data_collection}. Financial features for each stock/index
(Open, Close, Return, Trade Volume and Volatility) is correlated
with Twitter features (Positive, Negative, Bullishness, Carried Positive,
Carried Negative and Carried Bullishness).The time period under study
is monthly average as it the most accurate time window that gives
significant values as compared to other time windows which is
discussed later section~\ref{prediction_accuracy}.

Our approach shows strong correlation values between various features (upto $-0.96$ for opening price of Oracle and $0.88$ for returns from DJIA index etc.) and the average value of correlation between various features is around $0.5$.
Comparatively highest correlation values from earlier work has been around $0.41$ ~\cite{Sprenger}.
As the relationships between the stock(index) parameters and Twitter
features show different behavior in magnitude and sign for different
stocks(indices), a uniform standardized model would not applicable
to all the stocks(indices). Therefore, building an individual model
for each stock(index) is the correct approach for finding appreciable
insight into the prediction techniques.
Trading volume is mostly governed by agreement values of tweet feeds
as $-0.7$ for same day agreement and $-0.65$ for DJIA. Returns are
mostly correlated to same day bullishness by $0.61$ and by lesser
magnitude $0.6$ for the carried bullishness for DJIA. Volatility is
again dependent on most of the Twitter features, as high as $0.77$
for same day message volume for NASDAQ-100.

One of the {\it anomalies} that we have observed is that EBay
gives negative correlation between the all the features due
to heavy product based marketing on Twitter which turns out
as not a correct indicator of average growth returns of the
company itself.

\begin{table*}[htll]
  \centering
  \caption{Granger's Casuality Analysis of DJIA and NASDAQ for 7 week lag Twitter sentiments. (NSDQ is short for NASDAQ)}
  \scriptsize
    \addtolength{\tabcolsep}{-5.5pt}
    \centering
    \begin{tabular}{|p{1cm}|p{0.5cm}|p{1cm}|p{1.2cm}|p{1cm}|p{1cm}|p{1cm}|p{1cm}|p{1cm}|p{1cm}|p {1.15cm}|p{1cm}|}
    \toprule
    \textbf{Index $\Downarrow$} & \textbf{Lag} & \textbf{Positive} & \textbf{Negative} & \textbf{ Bull} & \textbf{ Agrmnt} & \textbf{Msg Vol} & \textbf{Carr Positive} & \textbf{Carr Negative} & \textbf{ Carr Bull} & \textbf{Carr Agrmnt} & \textbf{ Carr Msg Vol} \\ \hline
    \midrule
    \multicolumn{1}{|c}{\multirow{7}[0]{*}{{\textbf{DJIA}}}} & 1 & 0.614 & 0.122 & 0.891 & 0.316 & 0.765 & 0.69  & 0.103 & 0.785 & 0.759 & 0.934 \\
    \multicolumn{1}{|c}{} & 2 & 0.033** & 0.307 & 0.037** & 0.094* & 0.086** & 0.032** & 0.301** & 0.047** & 0.265 & 0.045** \\
    \multicolumn{1}{|c}{} & 3 & 0.219 & 0.909 & 0.718 & 0.508 & 0.237 & 0.016** & 0.845 & 0.635 & 0.357 & 0.219 \\
    \multicolumn{1}{|c}{} & 4 & 0.353 & 0.551 & 0.657 & 0.743 & 0.743 & 0.116 & 0.221 & 0.357 & 0.999 & 0.272 \\
    \multicolumn{1}{|c}{} & 5 & 0.732 & 0.066 & 0.651 & 0.553 & 0.562 & 0.334 & 0.045** & 0.394 & 0.987 & 0.607 \\
    \multicolumn{1}{|c}{} & 6 & 0.825 & 0.705 & 0.928 & 0.554 & 0.732 & 0.961 & 0.432 & 0.764 & 0.261 & 0.832 \\
    \multicolumn{1}{|c}{} & 7 & 0.759 & 0.581 & 0.809 & 0.687 & 0.807 & 0.867 & 0.631 & 0.987 & 0.865 & 0.969 \\ \hline
     \hline \multicolumn{1}{|c}{\multirow{7}[0]{*}{{\textbf{NSDQ}}}} & 1 & 0.106 & 0.12  & 0.044** & 0.827 & 0.064* & 0.02** & 0.04** & 0.043** & 0.704 & 0.071* \\
    \multicolumn{1}{|c}{} & 2 & 0.048** & 0.219 & 0.893 & 0.642 & 0.022** & 0.001** & 0.108 & 0.828 & 0.255 & 0.001** \\
    \multicolumn{1}{|c}{} & 3 & 0.06* & 0.685 & 0.367 & 0.357 & 0.135 & 0.01** & 0.123 & 0.401 & 0.008** & 0.131 \\
    \multicolumn{1}{|c}{} & 4 & 0.104 & 0.545 & 0.572 & 0.764 & 0.092* & 0.194 & 0.778 & 0.649 & 0.464 & 0.343 \\
    \multicolumn{1}{|c}{} & 5 & 0.413 & 0.997 & 0.645 & 0.861 & 0.18  & 0.157 & 0.762 & 0.485 & 0.945 & 0.028 \\
    \multicolumn{1}{|c}{} & 6 & 0.587 & 0.321 & 0.421 & 0.954 & 0.613 & 0.795 & 0.512 & 0.898 & 0.834 & 0.591 \\
    \multicolumn{1}{|c}{} & 7 & 0.119 & 0.645 & 0.089 & 0.551 & 0.096 & 0.382 & 0.788 & 0.196 & 0.648 & 0.544 \\
    \bottomrule
    \end{tabular}%
  \label{tab:grangers}%
\end{table*}%

\subsection{Bivariate Granger Causality Analysis}
\label{GCA}

The results in previous section show strong correlation
between financial market parameters and Twitter sentiments.
However, the results also raise a point of discussion:
Whether market movements affects Twitter sentiments or Twitter
features causes changes in the markets? To verify this hypothesis
we make use of Granger Causality Analysis (GCA) to the time series
averaged to weekly time window to returns through DJIA and
NASDAQ-100 with the Twitter features (positive, negative,
bullishness, message volume and agreement). GCA is not used to establish causality, but as an economist tool to investigate a statistical pattern of lagged correlation. A similar observation that cloud precede rain is widely accepted; proving cloud may may contain something that causes rain but itself may not be actual causative of the real event.

GCA rests on the assumption that if a variable X causes Y then changes in X
will be systematically occur before the changes in Y. We realize
lagged values of X shall bear significant correlation with Y.
However correlation is not necessarily behind causation. We have
made use of GCA in similar fashion as ~\cite{Bollen_Mao_Zeng_2010,Gilbert_Karahalios_2010}
This is to test if one time series is significant in predicting another time
series. Let returns $R_t$ be reflective of fast movements in the
stock market. To verify the change in returns with the change in
Twitter features we compare the variance given by following linear
models in equation 5 and equation 6 -

\begin{equation}
\vspace{-0.1in}
R_{t}= \alpha +\sum_{i=1}^{n}\beta_{i}D_{t-i}+ \epsilon_{t}
\vspace{-0.1in}
\end{equation}

\begin{equation}
R_t= \alpha+\sum_{i=1}^{n}\beta_iD_{t-i}+ \sum_{i=1}^{n}{\gamma_i X_{t-i}}+ \varepsilon_t
\end{equation}

Equation 5 uses only '$n$' lagged values of $R_t$ , i.e. ($R_{t-1},
. . ., R_{t-n}$ ) for prediction, while Equation 6 uses the
$n$ lagged values of both $R_t$ and the tweet features time series
given by $X_{t-1}$, . . . , $X_{t-n}$. We have taken weekly time window
to validate the casuality performance, hence the lag values
\footnote{{\it lag at k} for any parameter M at $x_{t}$ week is the value
of the parameter prior to $x_{t-k}$ week. For example, value of returns for the month of
April, at the lag of one month will be $return_{april-1}$ which will
be $return_{march}$}will be calculated over the weekly intervals $1,2,...,7$.

From the Table ~\ref{tab:grangers}, we can reject the null hypothesis $(H_o)$
that {\it the Twitter features do not affect returns in the financial
markets} i.e. $\beta_{1,2,....,n} \neq 0$ with a high level of confidence (high p-values).
However as we see the result applies to only specific negative and
positive tweets (** for p-value $< 0.05$ and * for p-value $< 0.1$
which is 95\% and 99\% confidence interval respectively).
Other features like agreement and message volume do not have
significant casual relationship with the returns of a stock
index (low p-values).

\subsection{EMMS Model for Forecasting}
\label{EMMS}

We have used Expert Model Mining System (EMMS) which incorporates a set
of competing methods such as Exponential Smoothing (ES), Auto Regressive
Integrated Moving Average (ARIMA) and seasonal ARIMA models. These
methods are widely used in financial modeling to predict the values of
stocks/bonds/commodities/etc~\cite{Pegels,Box}. These methods are
suitable for constant level, additive trend or multiplicative trend and
with either no seasonality, additive seasonality, or multiplicative
seasonality.

In this work, selection criterion for the EMMS is coefficient of
determination (R squared) which is square of the value of pearson-'r' of
fit values (from the EMMS model) and actual observed values. Mean
absolute percentage error (MAPE) and maximum absolute percentage error
(MaxAPE) are mean and maximum values of error (difference between fit
value and observed value in percentage). To show the performance of
tweet features in prediction model, we have applied the EMMS twice - first
with
tweets features as independent predictor events and second time without
them. This provides us with a quantitative comparison of improvement in
the prediction using tweet features.

ARIMA (p,d,q) are in theory and practice, the most general class of
models for forecasting a time series data, which is subsequently
stationarized by series of transformation such as differencing or
logging of the series $Y_i$. For a non-seasonal ARIMA (p,d,q) model- p is
autoregressive term, d is number of non-seasonal differences and q is
the number of lagged forecast errors in the predictive equation.  A
stationary time series $\Delta Y$ differences $d$ times has stochastic
component

\begin{equation}
    \Delta Y_i \sim Normal(\mu_i,\sigma^2)
\end{equation}

Where $\mu_i$ and $\sigma^2$   are the mean and variance of normal
distribution, respectively. The systematic component is modeled as:

\begin{equation}
\vspace{-0.1in}
    \begin{split} \mu_i= \alpha_i\Delta Y_{i-1}+.....+
    \alpha_p\Delta Y_{i-p}+ \theta_i\varepsilon_{i-1}\\+.....+ \theta_i
    \varepsilon_{i-q} \end{split}
\end{equation}

Where, $\Delta Y$ the lag-p observations from the stationary time series
with associated parameter vector $\alpha$ and $\epsilon_i$ the lagged
errors of order q, with associated parameter vector. The expected value
is the mean of simulations from the stochastic component,

\begin{equation}
\begin {split} E(Y_{i})=\mu_i= \alpha_i\Delta
Y_{i-1}+.....+ \alpha_p\Delta Y_{i-p}+ \theta_i\varepsilon_{i-1}
\\ +.....+ \theta_i\varepsilon_{i-q} \end {split}
\end{equation}

\begin{table}[htbp]
  \centering
    \scriptsize
  \addtolength{\tabcolsep}{-3pt}
  \caption{EMMS model fit characteristics for DJIA and NASDAQ-100}
    \begin{tabular}{|c|c|ccc|ccc|}
    \toprule
    \multirow{2}[0]{*}{\textbf{Index}} & \multirow{2}[0]{*}{\textbf{Predictors}} & \multicolumn{3}{|c|}{\textbf{Model Fit statistics}} & \multicolumn{3}{|c|}{\textbf{Ljung-Box Q(18)}} \\ \cline{3-8}

       &    & \textbf{R-squared} & \textbf{MaxAPE} & \textbf{Direction} & \textbf{Statistics} & \textbf{DF} & \textbf{Sig.} \\ \midrule \hline
    \multirow{2}[0]{*}{\textbf{Dow-30}} & \textbf{Yes} & 0.95 & 1.76 & 90.8 & 11.36 & 18 & 0.88 \\
       & \textbf{No} & 0.92& 2.37 & 60 & 9.9 & 18 & 0.94 \\ \hline
    \multirow{2}[0]{*}{\textbf{NASDAQ-100}} & \textbf{Yes} & 0.68 & 2.69 & 82.8 & 23.33 & 18 & 0.18 \\
       & \textbf{No} & 0.65 & 2.94 & 55.8 & 16.93 & 17 & 0.46 \\ \hline
    \bottomrule
    \end{tabular}%
  \label{tab:Arima2}%
\end{table}%

Seasonal ARIMA model is of form ARIMA (p ,d ,q) (P,D,Q) where P
specifies the seasonal autoregressive order, D is the seasonal differencing order
and Q is the moving average order. Another advantage of EMMS model is
that it automatically selects the most significant predictors among all
others that are available.

In the dataset we have time series for a total of approximately 60 weeks (422 days), out of which we use approximately 75\% i.e. 45 weeks for the training both the models with and without the predictors for the time period June 2nd 2010 to April 14th 2011. Further we verify the model performance as one step ahead forecast over the testing period of 15 weeks from April 15th to 29th July 2011 which count for wide and robust range of market conditions. Forecasting accuracy in the testing period is compared for both the models in each case in terms of maximum absulute percentage error (MaxAPE), mean absolute percentage error (MAPE) and the direction accuracy. MAPE is given by the equation \ref{eqn:mape}, where $\hat{y_i}$ is the predicted value and $y_i$ is the actual value.

\begin{equation}
\label{eqn:mape}
 MAPE= \frac{{{\Sigma^{n}}_{i}} |\frac{y_i - \hat{y_i}}{y_i}|}{n} \times 100
\end{equation}

While direction accuracy is measure of how accurately market or commodity up/ down movement is predicted by the model, which is technically defined as logical values for $(y_{i, \hat{t}+1}- y_{i,t})\times(y_{i,t+1}- y_{i,t})>0$ respectively.

As we can see in the Table~\ref{tab:Arima2}, there is significant
reduction in MaxAPE for DJIA(2.37 to 1.76) and NASDAQ-100 (2.96 to 2.69)
when EMMS model is used with predictors as events which in our case our
all the Tweet features (positive, negative, bullishness, message volume
and agreement). Using tweet features as part of the prediction process
in the EMMS model, gives more robust approach than the traditional
forecasting methods. There is significant decrease in the value of MAPE
for DJIA which is $0.8$ in our case than $1.79$ for earlier
approaches~\cite{Bollen_Mao_Zeng_2010}. As we can from the
values of R-square, MAPE and MaxAPE in Table~\ref{tab:Arima2}
for both DJIA and NASDAQ $100$, our proposed model uses Twitter sentiment
analysis for a superior performance over traditional methods. Since EMMS
is a customizable and scalable technique, our proposed model is bound to
perform well in a wide range of stocks and indices.

\begin {figure}[h]
\centering
\includegraphics [scale=0.35]{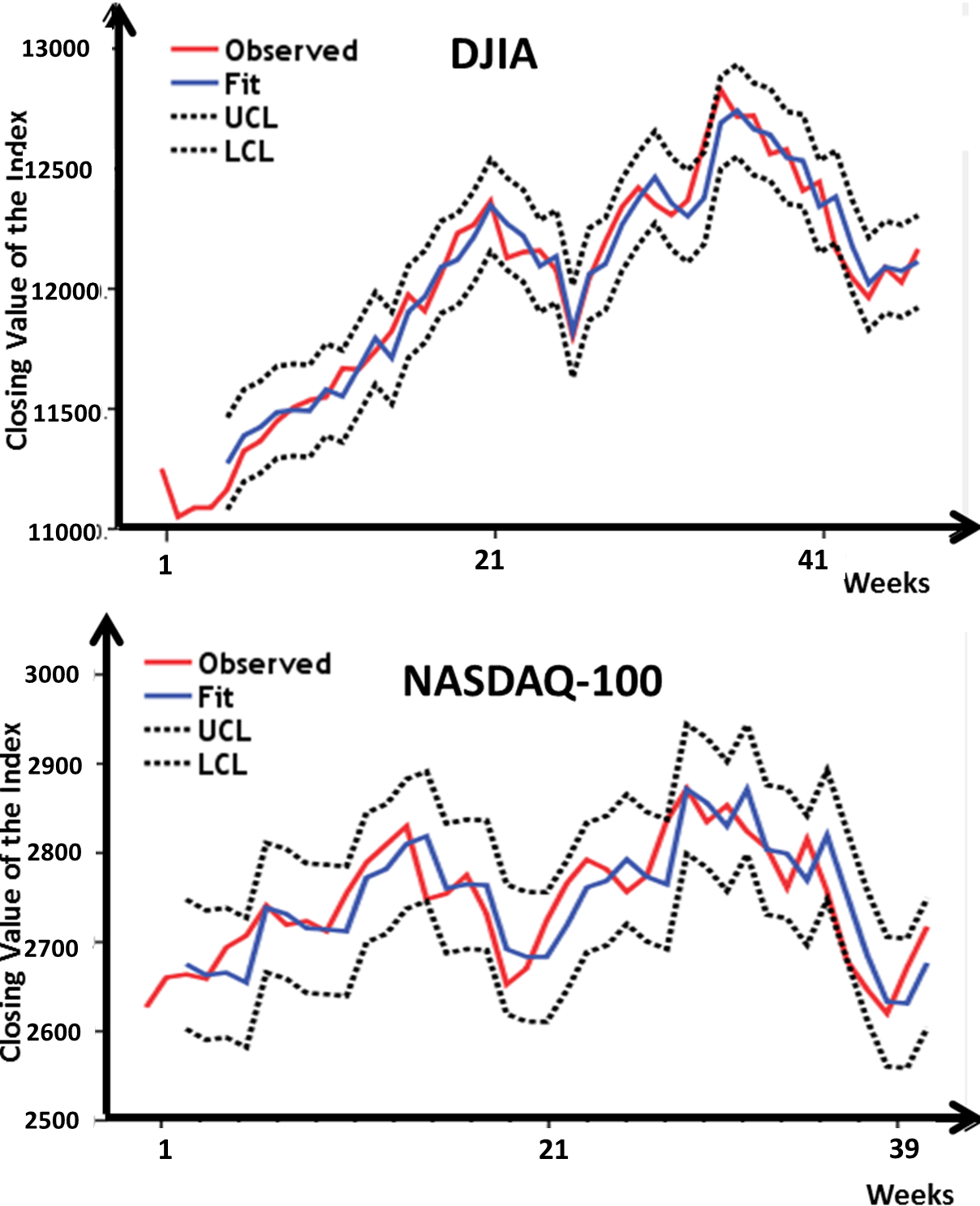}
\caption {Plot of Fit values (from the EMMS model) and actual observed closing values for DJIA and NASDAQ-100. Fit are the modeled values through EMMS, observed are the actual values of the index and UCL \& LCL are the upper and the lower confidence limit.}
\label{fig:Arima_dow}
\end{figure}

Figures ~\ref{fig:Arima_dow} shows the EMMS model fit for weekly closing
values for DJIA and NASDAQ $100$. In the figure fit are model fit values,
observed are values of actual index and UCL \& LCL are upper and lower
confidence limits of the prediction model.

\subsection{Prediction Accuracy using OLS Regression}
\label{prediction_accuracy}

\begin {figure}[h!]
\centering
\includegraphics [scale=0.55] {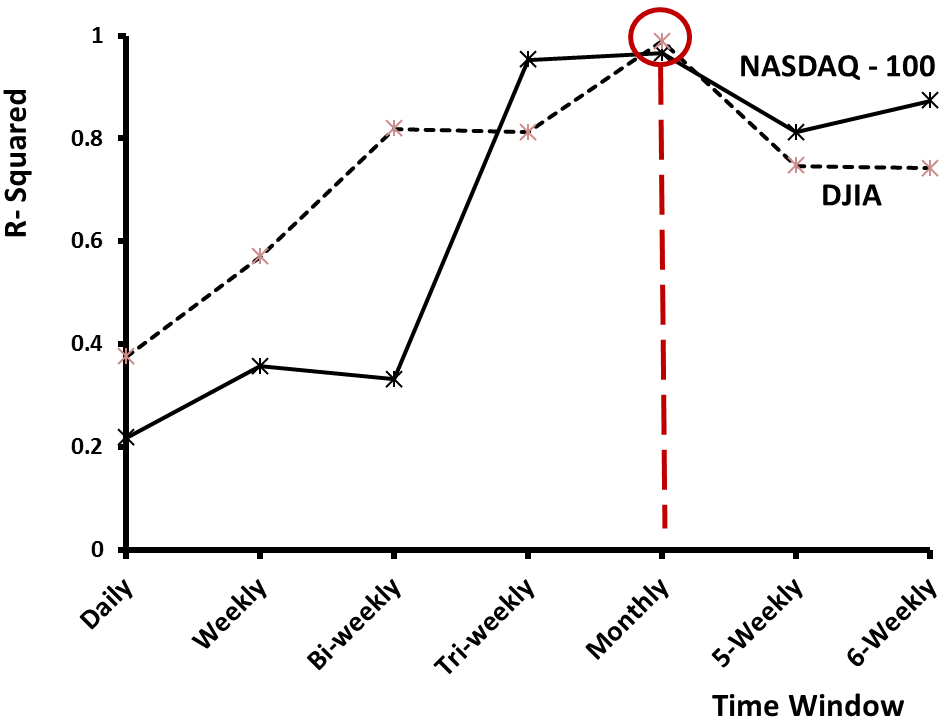}
\caption {Plot of R-square values over different time windows for DJIA and NASDAQ-100. Higher values
denote greater prediction accuracy.}
\label{fig:Ease_prediction}
\end{figure}

Our results in the previous section showed that forecasting
performance of stocks/indices using Twitter sentiments varies
for different time windows. Hence it is important to
quantitatively deduce a suitable time window that will give us most
accurate prediction. Figure~\ref{fig:Ease_prediction} shows
the plot of R-square metric for OLS regression for returns from stock
indexes NASDAQ-100 and DJIA from tweet board features (like number of
positive, negative, bullishness, agreement and message volume) both for
carried (at 1-day lag) and same week.

The R-square metric (explained in section~\ref{EMMS}) is calculated
as prediction performance indicator for different time
windows from daily, weekly, bi-weekly to 6 weekly time window.
From the figure ~\ref{fig:Ease_prediction} it can be inferred as we increase the time
window the accuracy in prediction increases but only till a certain
point that is monthly in our case beyond which value of R-square starts
decreasing again. {\it Thus, for monthly predictions we have highest
accuracy in predicting anomalies in the returns from the tweet board
features}.

In the next section we will discuss the practical implementation of how short term hedging strategies can improve efficiency by modeling mass public opinion and behavior for a particular company or stock index through mining of tweet sentiments.

\section{HEDGING STRATEGY USING TWITTER SENTIMENT ANALYSIS}
\label{hedging_startgies}

Portfolio protection is very important practice that is weighted as much as portfolio appreciation. Just like a normal user purchases insurance for its house, car or any commodity, one can also buy insurance for the investment that is made in the stock securities. This doesn't prevent a negative event from happening, but if it does happen and you're properly hedged, the impact of the event is reduced. In a diverse portfolio hedging against investment risk means strategically using instruments in the market to offset the risk of any adverse price movements. Technically, to hedge investor invests in two securities with negative correlations, which again in itself is time varying dynamic statistics.

To explain how weekly forecast based on mass tweet sentiment features can be potentially useful for a singular investor, we will take help of a simple example.

 Let us assume that the share for a company C1 is available for \$X per share and the cost of premium for a stock option of company C1 (with strike price \$X) is  \$Y.

 A $=$ total amount invested in shares of a company \emph{C1} which is number of shares (let it be N) $\times$ \$X

 B= total amount invested in put option of company \emph{C1} (relevant blocksize $\times$ \$Y)

 And always for an effective investment  (N $\times$ \$X) $>$  ( Blocksize $\times$ \$Y)

\begin {figure}[h!]
\centering
\includegraphics [scale=0.605] {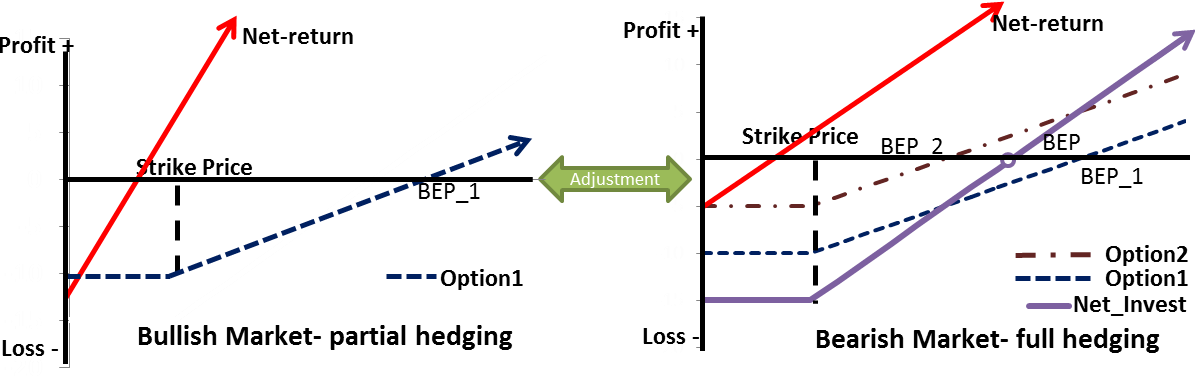}
\caption {Portfolio adjustment in cases of bearish (fully hedged) and bullish (partial hedged) market scenarios. In both the figures, strike price is the price at which a option is purchased, Break even point (BEP) is the instance when investment starts making profit. In case of bearish market scenario, two options at same strike price (but different premiums) are in purchased at different instances, Option1 brought at the time of initial investment and Option2 brought at a later stage (hence lower in premium value).}
\label{fig:Hedge_adjust_diag}
\end{figure}

An investor shall choose the value of N as per as their risk appetitive  i.e. ratio of A:B $=$ 2:1 (assumed in our example, will vary from from investor to investor). Which means in the rising market conditions, he would like to keep 50\% of his investment to be completely guarded, while rest 50\% are risky components; whereas in the bearish market condition he would like to keep his complete investment fully hedged by buying put options equivalent of all the investment he has made in shares for the same security. From Figure \ref{fig:Hedge_adjust_diag}, we infer for the P/L curves consisting of shares and 2 different put options for the company \emph{C1} purchased as different time intervals \footnote{The reason behind purchase of long put options at different time intervals is because in a fully hedged portfolio, profit arrow has lower slope as compared to partially hedged portfolio (refer P/L graph). Thus the trade off between risk and security has to be carefully played keeping in mind the precise market conditions.}; hence the different premium price even with the same strike price of \$X. Using married put strategy makes the investment risk free but reduces the rate of return in contrast to the case which comprises of only equity security which is completely free-fall to the market risk. Hence the success of married put strategy depends greatly on the accuracy of predicting whether the markets will rise of fall. Our proposed Tweet sentiment analysis can be highly effective in this prediction to determine accurate instances when the investor should readjust his portfolio before the actual changes happen in the market. Our proposed approach provides an innovative technique of using dynamic Twitter sentiment analysis to\textit{exploit the collective wisdom of the crowd for minimising the risk in a hedged portfolio}. Below we summarize two different portfolio states at different market conditions.

\begin{table}[htb]
  \centering
  \caption{Example investment breakdown in the two cases}
    \begin{tabular}{|p{11.6cm}|}
     \hline \hline
    \textbf{Partially Hedged Portfolio at 50\% risk}\\ \hline
  1000 shares at price of \$X  $=$ 1000X\\
  1 Block size of 500 shares put options purchased at strike price of \$X with premium of \$Y each $=$ 500Y\\
  Total= 1000X + 500Y  \\ \hline \hline
   \textbf{Fully Hedged Portfolio at minimized risk}  \\ \hline
   1000 shares at price of \$X  $=$ 1000X\\
   2 Block size of 500 shares each put options purchased at strike price of \$X with premium of \$Y each $=$ 2$\times$500Y $=$ 1000Y \\
  Total $=$ 1000X $+$ 1000Y \\ \hline \hline
    \end{tabular}%
  \label{tab:Hedging_descp}%
\end{table}%

\begin {figure}[htll]
\centering
\includegraphics [scale=0.35]{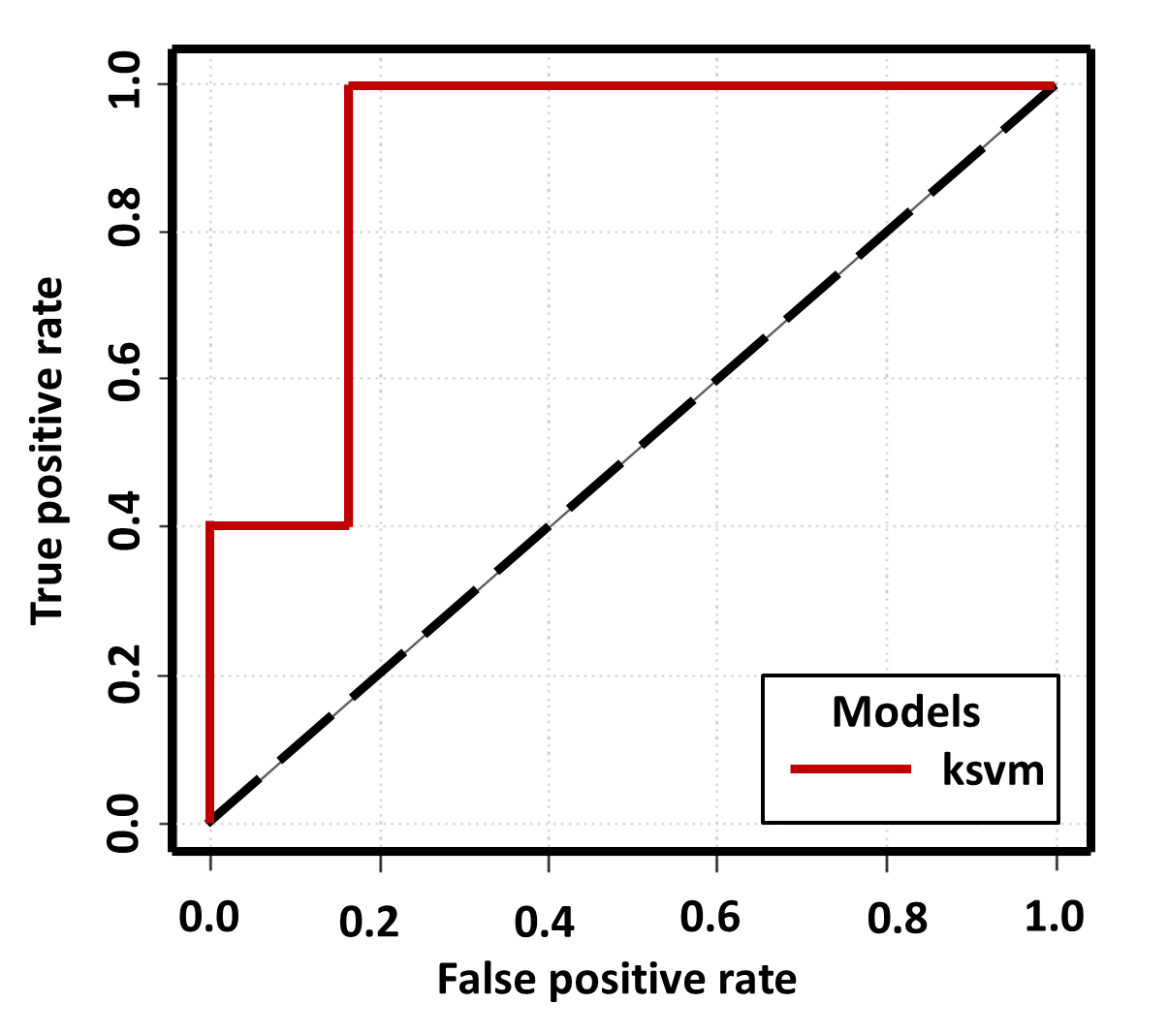}
\caption {Receiver operating characteristic (ROC curve) curve for the KSVM classifier prediction over the testing phase. ROC is graphical plot of the sensitivity or true positive rate, vs. false positive rate (one minus the specificity or true negative rate). More the area under curve for typical ROC, more is the performance efficiency of the machine learning algorithm.}
\label{fig:roc_curve}
\end{figure}

To check the effectiveness of our proposed tweet based hedging strategy, we run simulations and make portfolio adjustments in various market conditions (bullish, bearish, volatile etc). To elaborate, we take an example of DJIA ETF's as the underlying security over the time period of 14th November 2010 to 30th June 2011. Approximately 76\% of the time period is taken in the training phase to tune the SVM classifier (using tweet sentiment features from the prior week). This trained SVM classifier is then used to predict market direction (DJIA's index movement) in the coming week. Testing phase for the classification model (class 1- bullish market $\uparrow$ and class 0- bearish market $\downarrow$) is from 8th May to 30th June 2011 consisting a total of 11 weeks.
SVM model is build using KSVM classification technique with the linear (vanilladot) kernel using the package 'e1071' in R statistical language. Over the training dataset, the tuned value of the objective function is obtained as $-4.24$ and the number of support vectors is $8$. Confusion matrix for the predicted over the actual values (in percentage) is given in Table \ref{tab:DJIA_confusion_matrix}. Overall classifier accuracy over the testing phase is \textit{$90.91\%$}. Receiver operator characteristics (ROC) curve measuring the accuracy of the classifier as true positive rate to false positive rate is given in the figure \ref{fig:roc_curve}. It shows the tradeoff between sensitivity i.e. true positive rate  and specificity i.e. true negative rate (any increase in sensitivity will be accompanied by a decrease in specificity). Good statistical significance for the classification accuracy can be inferred from the value of area under the ROC curve (AUC) which comes out to $0.88$.

\begin{table}[htb]
  \centering
  \caption{Prediction accuracy over the Testing phase (11 weeks). Values in percentage.}
    \begin{tabular}{|c|c|cc|}
    \toprule
    \multicolumn{2}{|c|}{\multirow{2}[0]{*}{Confusion Matrix}} & \multicolumn{2}{|c|}{Predicted Direction}    \\  \cline{3-4}
    \multicolumn{2}{|c|}{} & Market Down & Market Up \\  \midrule  \hline
    \multirow{2}[0]{*}{Actual Direction} & Market Down & 45 & 9 \\
       & Market Up & 0  & 45 \\ \hline
    \bottomrule
    \end{tabular}%
  \label{tab:DJIA_confusion_matrix}%
\end{table}%

\begin {figure}[htll]
\centering
\includegraphics [scale=0.65]{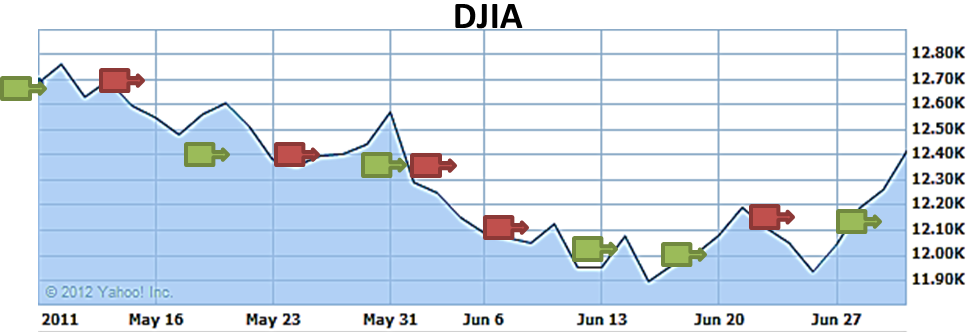}
\caption {DJIA index during the testing period. In the figure green marker shows adjustment bearish $\longrightarrow $ bullish, while red arrow shows adjustment bullish $\longrightarrow $ bearish. (Data courtesy Yahoo! finance)}
\label{fig:djia_index}
\end{figure}

Figure \ref{fig:djia_index} shows the DJIA index during the testing period and the arrows mark the weeks when the adjustment is done in the portfolio based on prediction obtained from tweet sentiment analysis of prior week. At the end of the week (on Sunday), using tweet sentiment feature we predict what shall be the market condition in the coming week- whether the prices will go down or up. Based on the prediction portfolio adjustment - bearish $\longrightarrow$ bullish or bullish $\longrightarrow$ bearish.

\section{DISCUSSIONS}
\label{discussions}
In section \ref{results}, we observed how the statistical behavior of market
through Twitter sentiment analysis provides dynamic window to
the investor behavior. Furthermore, in the section \ref{hedging_startgies}
we discussed how behavioral finance can be exploited in portfolio
decisions to make highly reduced risked investment. Our work answers
the important question - If someone is talking bad/good about
a company (say Apple etc.) as singular sentiment irrespective of
the overall market movement, is it going to adversely affect the
stock price? Among the 5 observed Twitter
message features both at same day and lagged intervals we realize
only some are Granger causative of the returns from DJIA and
NASDAQ-100 indexes, while changes in the public sentiment is well
reflected in the return series occurring at even lags of $1,2$ and
$3$ weeks. Remarkably the most significant result is obtained for
returns at lag 2 (which can be inferred as possible direction for
the stock/index movements in the next week).

Table~\ref{tab:Research_comparison} given below explains the different
approaches to the problem  that have been done in past by
researchers~\cite{Sprenger}, \cite{Bollen_Mao_Zeng_2010} and \cite{Gilbert_Karahalios_2010}.
As can be seen from the table, our approach is scalable, customizable and
verified over a large data set and time period as compared to other
approaches. Our results are significantly better than the previous
work. Furthermore, this model can be of effective use in formulating
short-term hedging strategies (using our proposed Twitter based
prediction model).

\begin{table*}[htll]
  \caption{Comparison of Various Approaches for Modeling Markets Movements Through Twitter}
  \addtolength{\tabcolsep}{-5pt}
\begin{center}
  {\scriptsize
  \begin{tabular}{|p{2.6cm}|p{2.9cm}|p{2.9cm}|p{3cm}|}
    \addlinespace
    \toprule
    Previous Approaches $\rightarrow$ & Bollen et al. \cite{Bollen_Mao_Zeng_2010} and Gilbert et al. \cite{Gilbert_Karahalios_2010} &  Sprenger et al. \cite{Sprenger} & This Work\\ \hline
    \midrule
    Approach & Mood of complete Twitter feed & Stock Discussion with ticker \$ on Twitter & Discussion based tracking of Twitter sentiments\\ \hline
    Dataset & 28th Feb 2008 to 19th Dec 2008, 9M tweets sampled as 1.5\% of Twitter feed & 1st Jan 2010 to 30th June 2010- 0.24M tweets & 2nd June 2010 to 29th
    July 2011- 4M tweets through search API\\ \hline
    Techniques & SOFNN, Grangers and linear models & OLS Regression and Correlation & Corr, GCA, Expert Model Mining System (EMMS)\\ \hline
    Results &
     * 86.7\% directional accuracy for DJIA & * Max corr value of 0.41 for returns of S\&P 100 stocks & \begin{tabular} {p{2.6cm}} * High corr values (upto -0.96) for opening price\\ * Strong corr values (upto 0.88) for returns\\ * MaxAPE of 1.76\% for DJIA \\ * Directional accuracy of 90.8\% for DJIA \end{tabular}\\ \hline
    Feedback/ Drawbacks & Individual modeling for stocks not feasible & News not taken into account, very less tweet volumes & Comprehensive and customizable
    approach. Can be used for hedging in F\&O markets \\ \hline
    \bottomrule
    \end{tabular}%
}
\end{center}
\label{tab:Research_comparison}%
\end{table*}

\section {CONCLUSION}
\label{conclusion}

In this paper, we have worked upon identifying relationships
between Twitter based sentiment analysis of a particular company/index
and its short-term market performance using large scale collection
of tweet data. Our results show that negative and positive dimensions of
public mood carry improved power to track movements of individual stocks/indices.
We have also investigated various other features like how previous week
sentiment features control the next week's opening, closing value
of stock indexes for various tech companies and major index like DJIA and
NASDAQ-100. As compared to earlier approaches in the area which have been
limited to wholesome public mood and stock ticker constricted discussions,
we verify strong performance of our alternate model that captures mass
public sentiment towards a particular index or company in scalable fashion
and hence empower a singular investor to ideate coherent relative comparisons.
Our analysis of individual company stocks gave strong correlation values (upto 0.88 for returns)
with twitter sentiment features of that company.
Further we also discuss how Twitter sentiments bring wisdom of the crowd
to use by even a singular investor in the form of simplistic married put
hedging strategy. Using this technique trader can retain his portfolio
with minimum risk even during highly bullish/bearish market conditions.
It is no surprise that this approach is far more robust and gives far better
results (upto 91\% directional accuracy) than any previous work.
In the near future, Twitter sentiments analysis promises to be an
effective strategy for hedging the investments in the financial markets.

\bibliographystyle{plain}
\bibliography{bibliography}

\begin{thebibliography}{10}

\bibitem{Acemoglu2010194}
Daron Acemoglu, Asuman Ozdaglar, and Ali ParandehGheibi.
\newblock Spread of (mis)information in social networks.
\newblock {\em Games and Economic Behavior}, 70(2):194 -- 227, 2010.

\bibitem{Asur_Huberman_2010}
Sitaram Asur and Bernardo~A Huberman.
\newblock Predicting the future with social media.
\newblock {\em Computing}, 25(1):492–499, 2010.

\bibitem{Hathaway}
The Atlantic.
\newblock Does anne hathaway news drive berkshire hathaway's stock?, 2011.
\newblock This is an electronic document. Date of publication: [March 18 2011].
  Date retrieved: October 12, 2011. Date last modified: [Date unavailable].

\bibitem{Bagnoli199927}
Mark Bagnoli, Messod~D. Beneish, and Susan~G. Watts.
\newblock Whisper forecasts of quarterly earnings per share.
\newblock {\em Journal of Accounting and Economics}, 28(1):27 -- 50, 1999.

\bibitem{Bollen_Mao_Zeng_2010}
Johan Bollen, Huina Mao, and Xiao-Jun Zeng.
\newblock Twitter mood predicts the stock market.
\newblock {\em Computer}, 1010(3003v1):1--8, 2010.

\bibitem{Box}
George Edward~Pelham Box and Gwilym Jenkins.
\newblock {\em Time Series Analysis, Forecasting and Control}.
\newblock Holden-Day, Incorporated, 1990.

\bibitem{danah_article}
danah~m. boyd and Nicole~B. Ellison.
\newblock Social network sites: Definition, history, and scholarship.
\newblock {\em Journal of Computer-Mediated Communication}, 13(1):210--230,
  2007.

\bibitem{Brown:2002:SLI:560498}
John~Seely Brown and Paul Duguid.
\newblock {\em The Social Life of Information}.
\newblock Harvard Business School Press, Boston, MA, USA, 2002.

\bibitem{Da_Engelberg_Gao_2010}
Zhi Da, Joseph Engelberg, and Pengjie Gao.
\newblock In search of attention.
\newblock {\em Russell The Journal Of The Bertrand Russell Archives}, (919),
  2010.

\bibitem{Das_chen}
Sanjiv~R. Das and Mike~Y. Chen.
\newblock {Yahoo! for Amazon: Sentiment Parsing from Small Talk on the Web}.
\newblock {\em SSRN eLibrary}, 2001.

\bibitem{Dewally}
Michaël Dewally.
\newblock Internet investment advice: Investing with a rock of salt.
\newblock {\em Financial Analysts Journal}, 59(4):65--77, 2003.

\bibitem{tweet_disaster}
S.~{Doan}, B.-K.~H. {Vo}, and N.~{Collier}.
\newblock {An analysis of Twitter messages in the 2011 Tohoku Earthquake}.
\newblock {\em ArXiv e-prints}, September 2011.

\bibitem{Ant_frank}
Murray~Z. Frank and Werner Antweiler.
\newblock {Is All That Talk Just Noise? The Information Content of Internet
  Stock Message Boards}.
\newblock {\em SSRN eLibrary}, 2001.

\bibitem{Garman_Klass}
Mark~B. Garman and Michael~J. Klass.
\newblock On the estimation of security price volatilities from historical
  data.
\newblock {\em The Journal of Business}, 53(1):67--78, 1980.

\bibitem{Gilbert_Karahalios_2010}
Eric Gilbert and Karrie Karahalios.
\newblock Widespread worry and the stock market.
\newblock {\em Artificial Intelligence}, pages 58--65, 2010.

\bibitem{Alec_Bhayani}
Alec Go, Richa Bhayani, and Lei Huang.
\newblock {Twitter Sentiment Classification using Distant Supervision}.

\bibitem{Guresen201110389}
Erkam Guresen, Gulgun Kayakutlu, and Tugrul~U. Daim.
\newblock Using artificial neural network models in stock market index
  prediction.
\newblock {\em Expert Systems with Applications}, 38(8):10389 -- 10397, 2011.

\bibitem{Lee1999357}
K.H. Lee and G.S. Jo.
\newblock Expert system for predicting stock market timing using a candlestick
  chart.
\newblock {\em Expert Systems with Applications}, 16(4):357 -- 364, 1999.

\bibitem{lerman}
Alina Lerman.
\newblock {Individual Investors' Attention to Accounting Information: Message
  Board Discussions}.
\newblock {\em SSRN eLibrary}, 2011.

\bibitem{morning_paper}
Huaxia~Rui Liangfei~Qiu and Andrew Whinston.
\newblock A twitter-based prediction market: Social network approach.
\newblock {\em ICIS 2011 Proceedings. Paper 5}, 2011.

\bibitem{Burton_m}
Burton~G. Malkiel.
\newblock The efficient market hypothesis and its critics.
\newblock {\em Journal of Economic Perspectives}, 17(1):59--82, 2003.

\bibitem{Bollen_second_paper}
Huina Mao, Scott Counts, and Johan Bollen.
\newblock Predicting financial markets: Comparing survey,news, twitter and
  search engine data.
\newblock Quantitative Finance Papers 1112.1051, arXiv.org, December 2011.

\bibitem{Pegels}
Garth~P. McCormick.
\newblock Communications to the editor—exponential forecasting: Some new
  variations.
\newblock {\em Management Science}, 15(5):311--320, 1969.

\bibitem{TIME09}
Douglas McIntyre.
\newblock Turning wall street on its head, 2009.
\newblock This is an electronic document. Date of publication: [May 29, 09].
  Date retrieved: September 24, 2011. Date last modified: [Date unavailable].

\bibitem{Hong}
Hong Miao, Sanjay Ramchander, and J.~K. Zumwalt.
\newblock {Information Driven Price Jumps and Trading Strategy: Evidence from
  Stock Index Futures}.
\newblock {\em SSRN eLibrary}, 2011.

\bibitem{BBC_twitter}
BBC News.
\newblock Twitter predicts future of stocks, 2011.
\newblock This is an electronic document. Date of publication: [April 6, 2011].
  Date retrieved: October 21, 2011. Date last modified: [Date unavailable].

\bibitem{Qian}
Bo~Qian and Khaled Rasheed.
\newblock Stock market prediction with multiple classifiers.
\newblock {\em Applied Intelligence}, 26:25--33, February 2007.

\bibitem{Sprenger}
Timm~O. Sprenger and Isabell~M. Welpe.
\newblock {Tweets and Trades: The Information Content of Stock Microblogs}.
\newblock {\em SSRN eLibrary}, 2010.

\bibitem{Szomszor_Kostkova_Quincey_2009}
Martin Szomszor, Patty Kostkova, and Ed~De Quincey.
\newblock swineflu : Twitter predicts swine flu outbreak in 2009.
\newblock {\em 3rd International ICST Conference on Electronic Healthcare for
  the 21st Century eHealth2010}, (December), 2009.

\bibitem{Tumasjan_Sprenger_Sandner_Welpe_2010}
Andranik Tumasjan, Timm~O Sprenger, Philipp~G Sandner, and Isabell~M Welpe.
\newblock Predicting elections with twitter: What 140 characters reveal about
  political sentiment.
\newblock {\em International AAAI Conference on Weblogs and Social Media
  Washington DC}, pages 178--185, 2010.

\bibitem{Wysocki_1998}
Peter Wysocki.
\newblock Cheap talk on the web: The determinants of postings on stock message
  boards.
\newblock {\em Working Paper}, 1998.

\bibitem{Business_week}
Max Zeledon.
\newblock Stocktwits may change how you trade, 2009.
\newblock This is an electronic document. Date of publication: [2009]. Date
  retrieved: September 01, 2011. Date last modified: [Date unavailable].

\bibitem{Zhang_Fuehres_Gloor_2009}
Xue Zhang, Hauke Fuehres, and Peter~A Gloor.
\newblock Predicting stock market indicators through twitter "i hope it is not
  as bad as i fear.
\newblock {\em Anxiety}, pages 1--8, 2009.

\end{thebibliography}
\newpage

\section{APPENDIX}
\vspace{-0.08cm}
Correlation heatmap indicative of significant relationships between various twitter features with the index features.
\vspace{-0.1cm}

\begin {figure}[h!]
\centering
\includegraphics [scale=0.6] {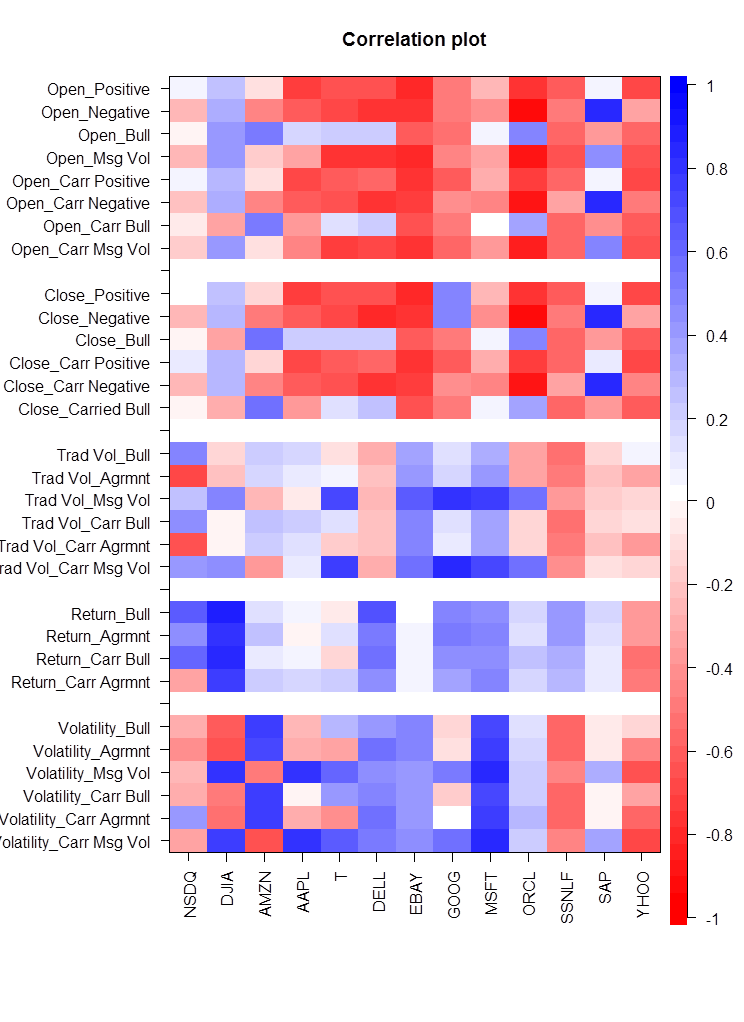}
\caption {Heatmap showing pearson correlation coefficients between security indices vs features from Twitter and stock features.}
\vspace{-0.1mm}
\label{fig:corr_heatmap}
\end{figure}
\vspace{-1mm}

\end{document}